\documentclass[aps,prl,twocolumn,showpacs,superscriptaddress,groupedaddress]{revtex4}

\usepackage{times,mathptmx,amssymb}
\usepackage{amsmath} 
\usepackage{subfigure}
\usepackage{epsfig}
\usepackage{graphicx}

\begin{document}

\title{Granular Segregation in Tapered Rotating Drums}

\author{S. Gonz\'alez}\thanks{\mbox{s.gonzalez@tugraz.at}}
\author{L. Orefice}
\affiliation{Research Center Pharmaceutical Engineering GmbH,
  Inffeldg. 13, Graz, Austria} 

\begin{abstract}
  We study the granular segregation in a tapered rotating drum by
  means of simulations. In this geometry, both radial and axial
  segregation appear, the strength of which depends on the filling
  fraction. We study the effect of the drum's tapering angle on the
  segregation speed and show that, coherently with several recent
  studies, the axial segregation is due to the combined effect of the
  radial segregation and the shape of the drum.  We show that the
  axial segregation behaves in a way analogous to the one found in
  chute flows, and corresponds to previous theories for shallow
  gravity driven surface flows with diffusion. By means of this
  analogy, we show that tapered drums could be the simplest
  experimental set-up to obtain the free parameters of the
  theory.
\end{abstract}

\maketitle 
 

It has been hinted that axial segregation in rotating drums is mainly
the result of the shape of the free-surface, the thin layer of flowing
particles in the drum
\cite{wouter2012,Pohlman2012,Zaman2013,Chen2011}. The free-surface is
not flat, but slightly curved along the axial direction, with peaks
near the parallel walls of the drum. Due to the friction with them,
the particles on the sides are lifted more than the particles in the
centre, thus creating a convective flow in the drum that induces the
axial segregation. How to harvest this relation between flow and the
segregation pattern remains an open question: a question whose answer
could have far-reaching practical implications.

Motivated by the results of \cite{Kawaguchi2006,Yada2010} -- where
axial segregation of binary mixtures in a tapered rotating drum is
reported albeit not explained -- and how they relate to our own results
\cite{Gonzalez2013} -- where the shape of the rotating drum is used to
speed up and control the axial segregation -- we study the segregation
pattern in tapered rotating drums by means of simulations, and see to
what extent it is possible to control it by modifying the drum's shape.

It has recently been shown that, contrary to the common belief that 3D
rotating drums are the sum of independent quasi-2D slices interacting
only via diffusion, the shape of the drum does influence the surface
flow creating a slow axial drift \cite{Zaman2013}. This drift is
caused by the axial slope of the drum's walls, which determines the 3D
structure of the flow. If a bi-disperse mixture is used, the particles
will follow the flow structure and will segregate accordingly.

In the particular case of a tapered drum, when mixed particles enter
the flowing layer in the upper part of the avalanche, gravity drives
them towards the small side of the drum before returning them to the
large one. Small particles can percolate trough the voids created by
the big ones leaving the flowing layer before it changes its
direction. This leads to a net transport of small particles to the
small side of the drum and of large particles to the big side: axial
segregation occurs. Since the shape of the flowing layer depends on
the shape of the drum, one would also expect it to affect the
segregation. In this paper we show that this is indeed the case and we
study how it happens.

The coupling of axial segregation with the velocity profile has been
already reported for the classical rotating drum \cite{Chen2011}. In
the tapered rotating drum the same coupling is in play but with a
fundamental difference: the shape of the drum increases the curvature
of the flow lines on the flowing layer thus transporting more small
particles than in a parallel drum, and in contrast to the latter, {\em
  in a definite direction}.

The structure of the paper is as follow. First, the drum geometry and
and simulation details are explained. Second, we compare our results
for half-filled drums with experiments to validate our model. Third,
to minimize the effect of radial segregation, a lower filling fraction
is used and total segregation is observed. The effect of the drum's
tapering angle on the speed of the segregation is also
studied. Finally, comparisons will be drawn with simulations of
mixtures in 2 and 3D chutes and with a theory for shallow gravity
driven surface flows with diffusion.

\begin{figure}[htp!]
  \centering
  \epsfig{file=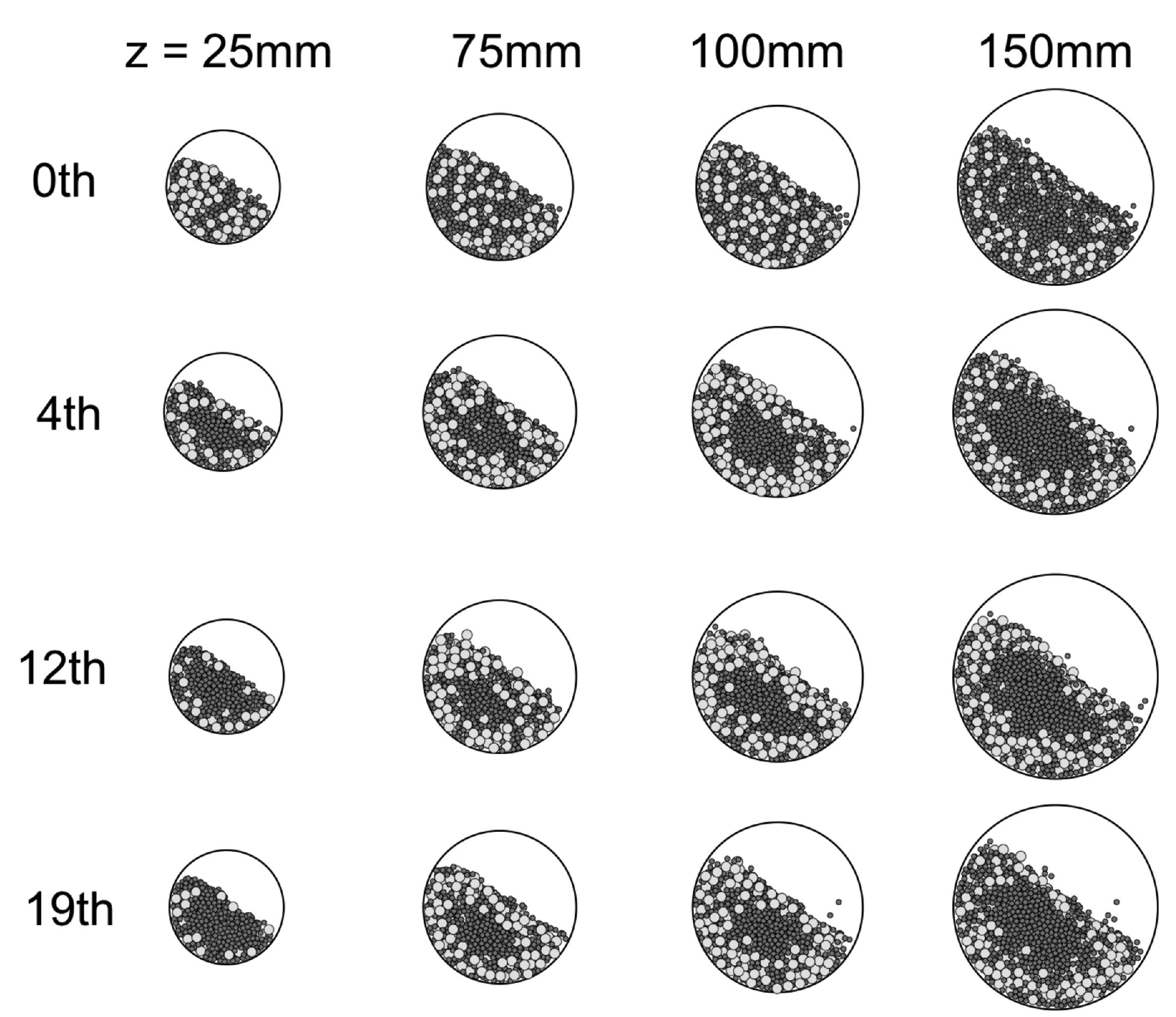,angle=0,width=\columnwidth} 
  \caption{(Color online) Snapshots for half-filled system at
    different positions along the rotation axis (from left to right)
    and different times (from top to bottom, the number of revolutions
    indicated in the left). Small particles dark, large particles
    light. One can see that first the radial segregation appears, and
    this slowly and continuously transforms into axial segregation,
    that here is visible in the dominance of small particle in the
    small side of the drum.}
\label{fig1}
\end{figure}

\section{Simulations}
Our simulations are in a slightly different geometry than that of
\cite{Kawaguchi2006,Yada2010} (shorter drum) to speed up the
simulation time. We use an equal volume mixture of 5 and 10mm radii
particles in a tapered drum 150mm length, with caps of radii $r = 50$
mm (at $z=0$mm) and variable R (at $z=150$mm), in the left and right
side, respectively. Their relation define the angle of the drum, such
that $R = r (1+\tan(\alpha))$. In the simulations, particles' contact
forces are represented using a standard spring dash-pot model
\cite{cundall79} as implemented in \cite{thornton2012}. The angular
velocity is set to $\pi\ s^{-1}$, i.e.\ 30rpm, which implies an
average Froude number of $Fr = \omega^2 (r+R)/2g \sim 0.1$. The system
is either at 50 or 25$\%$ filling fraction. As initial condition the
particles are arranged in a square lattice, alternating one big
particle and eight small to obtain an initial mixed state. Snapshots
of the system for different times and axial positions are shown in
Fig.\ \ref{fig1} (cf.\ Fig.\ 5 in Ref.\ \cite{Yada2010} for an
experimental comparison).

The simulations were performed using Mercury-DPM \cite{thornton2012}
with the same material properties used in our previous
research \cite{Gonzalez2013}, since they proved to reproduce the
experimental results.


\subsection{Shape of the Free Surface}
\begin{figure}[ht!]
  \centering

\epsfig{file=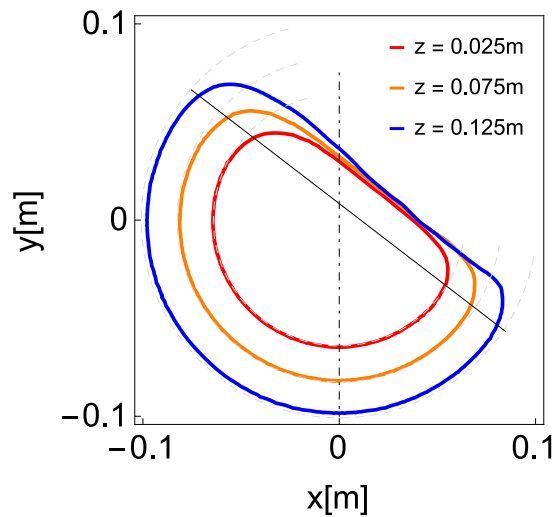,angle=90,width=.85\columnwidth,angle=-90}

  \caption{Contour lines for a density $\rho^* = \rho_{particles}/10$
    at three different axial positions for the same systems as in
    Fig.\ 1. The level difference can be seen mostly in the upper part
    of the avalanche. One should think of the curves as being
    displaced in the direction normal to the page. Compare the
    qualitative agreement with the model of Ref.\ \cite{zik1994}. Note
    that since the flow velocity increases along $z$, the blue curve
    shows a slightly more S-shaped profile. Solid line shows the plane
    where the velocity is measured for Fig.\ \ref{fig:velField} and
    dashed vertical line the plane where the density is measure for
    Fig.\ \ref{fig:packingFraction}.}
\label{fig:profile}
\end{figure}

\begin{figure*}[ht!]
  \centering

\epsfig{file=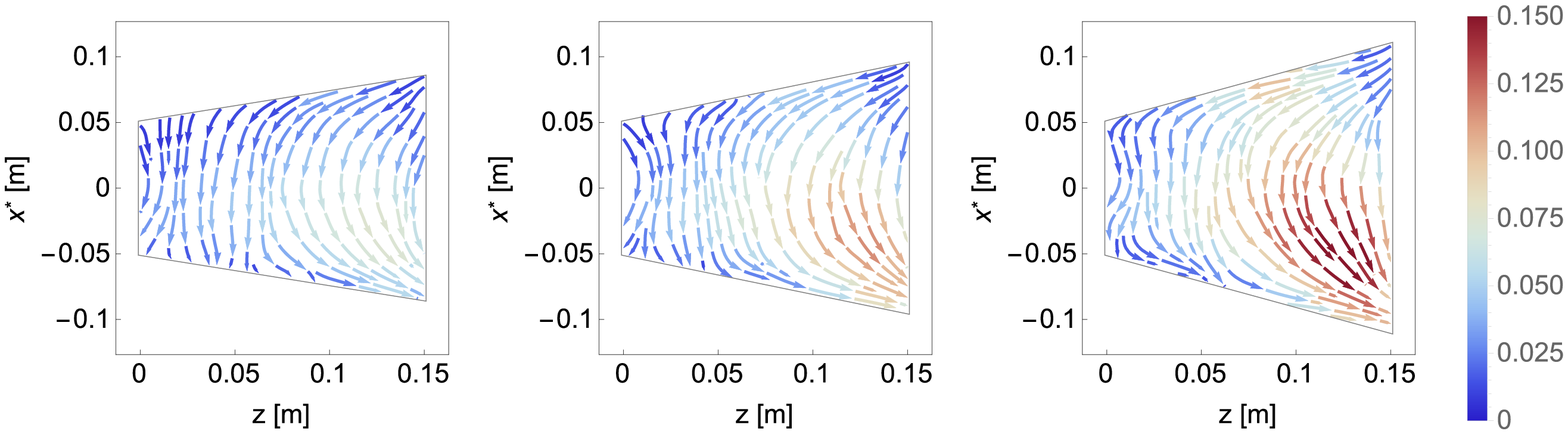,angle=0,width=1.875\columnwidth}
  \caption{Average velocity field of the particles' free surface for
    the tapered drum (from left to right, tapering angle
    $\alpha=10^\circ, 15^\circ, 20^\circ$), the colour representing the
    particles' mean velocity in $m/s$. The length of the component of the
    vector field parallel to the axis of rotation is increased to
    stress the effect of the tapered geometry on the flow. As
    expected, the convective flows increases its amplitude and reach
    with the angle of tapering. }
\label{fig:velField}
\end{figure*}

\begin{figure*}[ht!]
  \centering

\epsfig{file=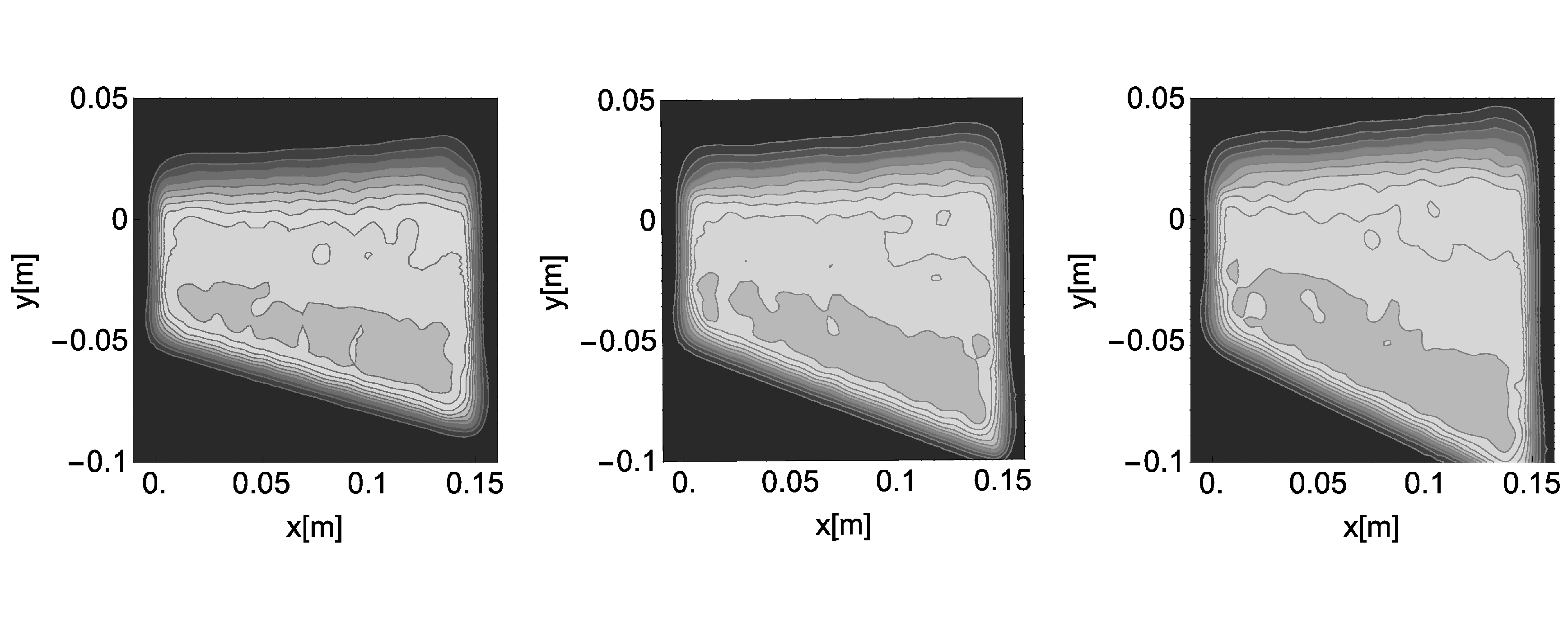,width=1.8\columnwidth}
\hfill \raisebox{.4cm}{
\epsfig{file=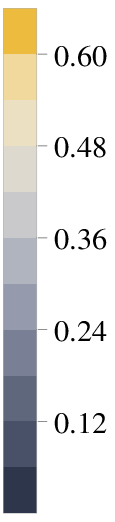,angle=0,width=.1\columnwidth}}
  \caption{(Color online) Packing fraction on the middle plane of the
    drum for increasing drum angles $\alpha =
    10^{\circ},15^{\circ},20^{\circ}$ , from left to right. The
    increasing slope of the free surface can be seen from left to
    right, as $\alpha$ increases. }
\label{fig:packingFraction}
\end{figure*}

In this section we show the three-dimensional structure of the flow.
We focus on three different characteristics: the level of the surface
along the axis of rotation (Fig. \ref{fig:profile}), the velocity
field on the free surface (Fig. \ref{fig:velField}) and the density
profile, as seen by a vertical cut in the centre of the drum
(Fig. \ref{fig:packingFraction}). Together, they provide a rather
complete way to visualise the structure. Data are averaged for 30
consecutive snapshots (one every 0.8s) and coarse-grained to obtain a
continuous function following the procedure of
Ref. \cite{Weinhart2012}. The weight function is a Lucy polynomial
\cite{Lucy1977} with a width of $2d$. Unless otherwise noted, every
coarse-graining presented here is done with the same parameters. We
measured at different times during the evolution of the system, and
the results do not significantly change, so we conclude that what we
observe depends only on the geometry and not in the degree of
segregation.

The free surface along the axial direction shows level differences.
This phenomenon, to the best of our knowledge, was first described by
Zik et al.\ for a drum modulated along the axial direction with a
sinusoidal function \cite{zik1994}. This can be seen in our
simulations from Fig.\ \ref{fig1}, or looking at the iso-lines of
the density at different axial positions, in Fig.\ \ref{fig:profile}. This
level difference along the axis makes particles to move along the
gradient of the surface, that is as particles go down the avalanche,
they move forwards and backwards in the axis perpendicular to the
page.

Particles enter the flowing layer with a small velocity and they
are mainly directed towards the small side of the drum. This is due to
the tapered geometry: when entering the flowing layer, the particles
in the large side of the drum are higher than those in the small side
of the drum, thus creating a flow towards the small side. This is
inverted when particles reach the lower part of the avalanche:
particles in the large side of the drum are lower than those in the
small side. Increasing the angle of the drum, increases the curvature
and the magnitude of this flow. This can be seen in
Fig.\ \ref{fig:velField} for three different angles. The surface is
defined as the plane whose projection can be seen in
Fig.\ \ref{fig:profile} (solid line).

When a mixture of particles enters the avalanche in the upper part,
they will segregate as they travel downslope. If the drum is large
enough, some small particles will reach the core before the flow
changes its direction, creating a net transport of small
particles. (Large enough here means in relation to the segregation
speed of the mixture. For size-driven segregation, the smaller the
size ratio of the particles, the slower segregation is, and thus the
drums needs to be larger.) The large particles, since they tend to
stay on the surface, can flow back to the large side of the drum,
following the flow lines. In this way, axial segregation occurs.

Finally, the packing fraction of the system gives us an insight on the
inner structure of the flow, and shows the dependence of the free
surface on the drum's angle, Fig.\ \ref{fig:packingFraction}. Despite
the presence of segregation, the density is uniform during the
evolution of the system, with the densest region at the bottom of the
drum. The density is mostly homogeneous along the axial direction with
increased spatial fluctuations as $\alpha$ increases. 

The density remains constants during the segregation process, and is
maximum in the lower region of the drum for all the angles studied. We
confirmed that the density does not vary by measuring it at different
times during the segregation process. 

\subsection{Radial and Axial Segregation}
To corroborate that the proposed mechanism is responsible of the axial
segregation, we measure the radial and axial segregation in the drum as
a function of time. In order to have order parameters for both, the
radial segregation is defined as follows:
\begin{equation}
  q_{\rm radial} = N_{s<z^*} /N_{Ts} +  N_{L>z^*} /N_{TL} - 1~,
\end{equation}
where $ N_{s<(L>)r^*}$ is the number of small (large) particles whose
radial coordinate is smaller (larger) than the half volume radius
$r^*$ (note that this radius depend on the axial coordinate),
$N_{Ts(L)}$ being the total number of small (large) particles. That
is, we count the fraction of particles that are in the expected side
of the drum. When $ q_{\rm radial} = 1$ all the large particles are in
the outside and the small in the core of the drum. It must be noted
that, due to the flowing layer, the radial segregation can never reach
the maximum value. To measure the axial segregation we use a quantity
that also works with banding (which appeared in one of the systems
studied) \cite{Taberlet2004}:

\begin{equation}
q_{axial}(t) = \frac{1}{l}\int_0^l\frac{|c_s(z,t) - \overline{c}_s(z)|}{ \overline{c}_s(z)}dz~,
\end{equation}
where $c_s(z,t)$ is the number of small particles along the axis, and
 $ \overline{c}_s(z)$ the number of small beads that correspond to a
slice of the drum at position $z$ for the totally mixed state. It must
be noted that since we are simulating a tapered rotating drum, the
mean number of small particles changes with $z$ since the cross
section of the drum increases accordingly.

\begin{figure}[h!]
  \centering
  \epsfig{file=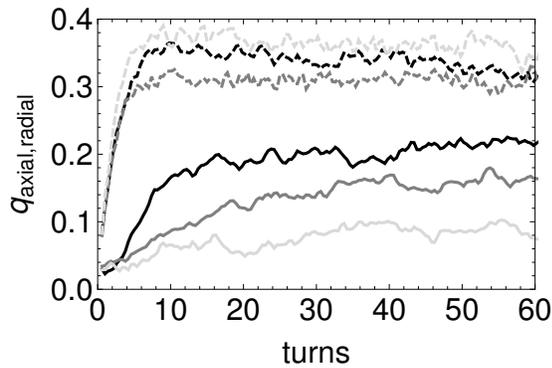,angle=0,width=.85\columnwidth}  

  \caption{Axial (solid-lines) and radial (dashed-lines) segregation
    as a function of time for the tapered rotating drum. From light
    gray to black in increasing angle of tapering. As it can be seen,
    there is no correlation between the angle and the radial
    segregation. However, axial segregation shows a clear correlation
    with $\alpha$.}
\label{fig:segtime}
\end{figure}

Figure \ref{fig:segtime} shows the temporal evolution for each kind of
segregation for three different angles. As expected, the radial
segregation sets in quite fast, essentially after four turns, but it
saturates soon to a roughly constant value. On the other hand, the
axial segregation is slower but steadily increases beyond the point of
saturation for the radial segregation. Eventually (around the 10th
revolution) the axial segregation causes the value of $q_{radial}$ to
decrease slightly, since in the small side of the drum the number of
large particles decreases greatly.

If $\alpha$ is increased the segregation speed should increase since
the level difference on the free surface is also increasing
\cite{zik1994}. This is indeed the case as \cite{Kawaguchi2006} shows
experimentally (see in particular Fig. 8). The same principle can be
seen in other configurations such as the spherical drum
\cite{Naji2009}, or the double cone configuration \cite{Albert2001}.

To better understand the role of geometry on the segregation process,
several simulations were run for different $\alpha$. It is expected
that the radial segregation will not depend on the geometry of
system. However, for the axial segregation, since it depends on the
velocity profile and on the geometry of the flowing layer, one expects
to find a strong dependence on the angle of the drum since this
determines the level difference in the flow. Indeed this is the case
as Fig.\ \ref{fig:segtime} shows.

\subsection{Less than Half Filled Drum}
\begin{figure}[ht!]
   \centering 
   \epsfig{file=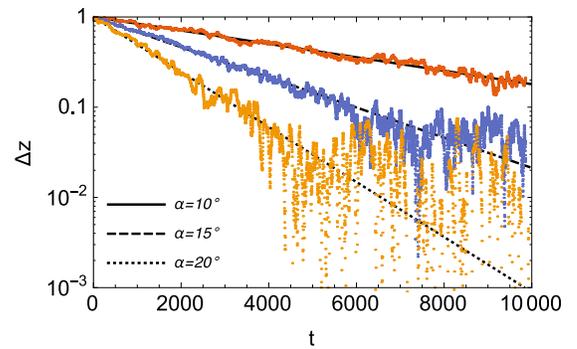,width=.85\columnwidth}

   \caption{Displacement of the centre of mass of large particles for
     systems with different $\alpha$, normalized by the asymptotic
     value $\Delta z$ in a log-linear scale. Straight lines correspond
     to the exponential fit. Compare with Ref.\ \cite{Staron2014}. If
     plotted in a linear scale, one obtains results as in
     Ref.\ \cite{Thornton2011}.}
 \label{fig:cmL}
 \end{figure}

Our previous result shed light on the relation between $\alpha$ and
the axial segregation, but this is strong influenced by the radial
segregation in half filled drum. So large in fact, that at this
filling fraction some systems segregate the other way around, that is,
large particles in the small side of the drum, as reported in
\cite{Kawaguchi2006} and confirmed by some of our simulations (data
not shown). We ran simulations at $25\%$ filling fraction, for
different $\alpha$, and we observe indeed that the axial segregation
is bolder. What is more, the radial segregation disappears almost
completely since the mixing at this filling fraction is maximum
\cite{Metcalfe95}. In this configuration, a pure phase of small
particles appears in the small side of the drum and a steady state for
the segregation is reached, while increasing the angle of the drum
increases the segregation speed (see Fig.\ \ref{fig:cmL}).

This behaviour is strikingly similar to that of bi-disperse chute
flows, either in 2D \cite{Staron2014} or 3D
\cite{Thornton2011}. Numerically, those are simulated with a box with
periodic boundary conditions in the flowing plane, and a rough
inclined bottom. Depending on the inclination of the bottom with
respect to gravity, a steady flow can be obtained. This is a
straightforward way to study segregation for a fixed velocity
profile, however their experimental realisation is rather complex.

Figure \ref{fig:cmL} shows the displacement of the centre of mass of
the large particles normalized by its asymptotic value, for three
different drum angles
\begin{equation}
\Delta z =  \frac{z_{cm}(\infty)  - z_{cm}(t)}{z_{cm}(\infty)  - z_{cm}(0)}~, 
\end{equation}
where $z_{cm}(t)$ is the centre of mass position on the axial
direction for the large particles. As $\alpha$ increases, the system
segregates faster since the shear rate increases with it. One can see
that the drum's angle influences the segregation in the same way as
the size ratio between the particles \cite{Thornton2011,marks2012} or
the volume fraction between the two species \cite{Staron2014}.

\begin{figure}[ht!]
   \centering 
   \epsfig{file=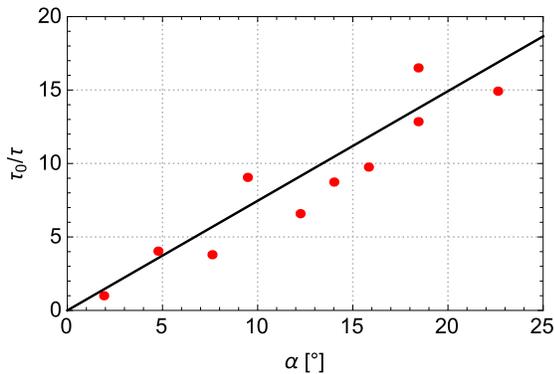,width=.85\columnwidth}
   \caption{Normalised segregation speed measured as the inverse of
     the typical time, $\tau_0/\tau$ obtained from the centre of mass
     displacement on Fig.\ \ref{fig:cmL} for different tapering angles
     ($\tau_0$ is the typical time for $\alpha=2^\circ$). The solid
     line is a linear eye guide.}
 \label{fig:speedAngle}
 \end{figure}

This is valid for a wide range of drum angles. We define the
segregation speed as the inverse of the exponent of the exponential
fit  $\Delta z = \exp(-t/\tau)$. In
general, the segregation speed is a monotonic function of the angle,
as one can see in Fig.\ \ref{fig:speedAngle}, the line being a linear
eye-guide. A simple estimation tells us that the distance a particle
travels on the free surface is $L_s \propto \tan(\alpha)$. If the
particles travel twice as much during one avalanche they will
segregate twice as fast, and this would explain the linear behaviour
of the segregation speed with the angle.

Since we have this qualitative agreement between two completely
different systems, the obvious next step is to see how well the
theoretical predictions developed for the chute flow work on the
tapered drum.

\subsection{Comparison with theory}

Let us look at the phenomenon from another point of view. Since the
free surface is at an angle with respect to the horizontal (going up
towards the large side of the drum, see Fig.\ 2) the projection of
gravity on the free surface is non-zero. This translates into an
effective gravity towards the small side of the drum. From the theory
for shallow granular avalanches \cite{gray05,gray11} we know the time
evolution for the concentration of each species, and we should be 
able to apply this result to the tapered drum.

Whereas the chute flow develops vertical segregation as it moves
downslope, the rotating drum develops axial segregation as time
passes. This is due to the structure of the flow as seen in
Fig.\ \ref{fig:velField}. The tapered drum turns the kinetic sieving
along the flow lines into axial segregation at the system's level.
According to this simple analogy, small particles should go to the
small side of the drum, while large particle will go towards the large
side. Furthermore, since the free-surface is a geometric
characteristic of the flow and does not depend on the degree of
segregation of the mixture, the tapered drum will segregate the
particles into stable configurations in the same manner that a chute
flow does.

\begin{figure*}[ht!]
  \centering

\hfill
\epsfig{file=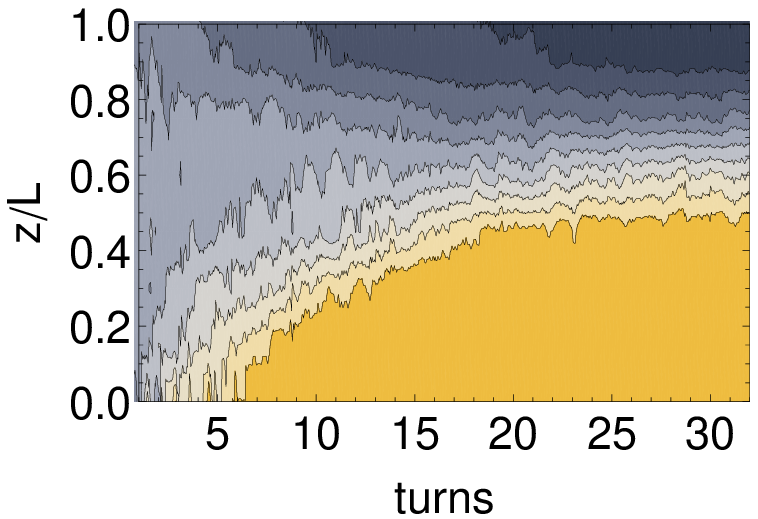,width=.85\columnwidth}
\hfill
\epsfig{file=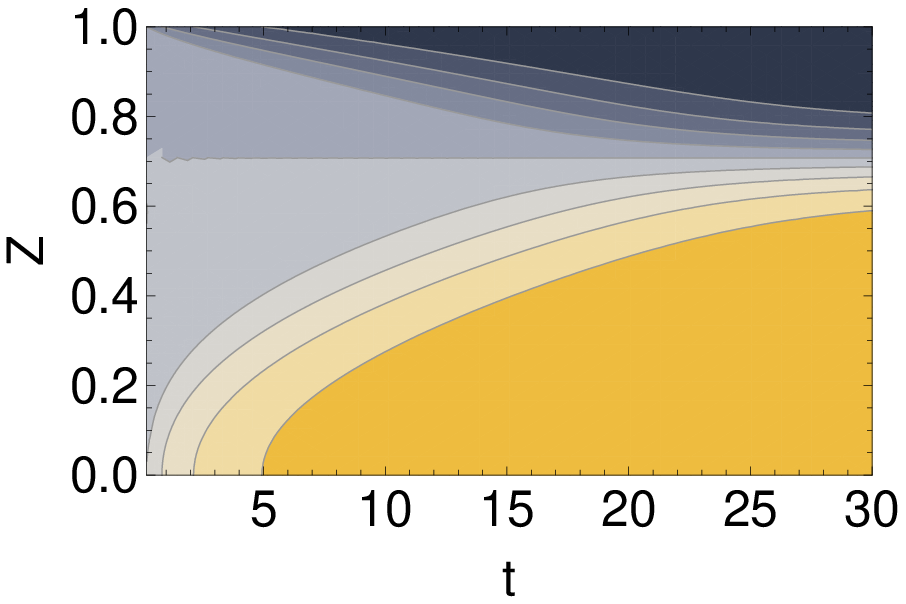,width=.875\columnwidth} \hspace{0.5cm}
\raisebox{0.25cm}{\epsfig{file=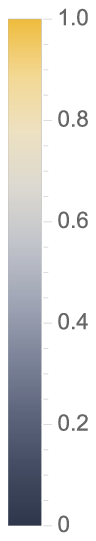,width=.12\columnwidth} }
   \caption{Left panel, concentration of small particles along the
     normalized rotation axis as a function of time for our numerical
     experiments with $\alpha=10^\circ$ and $L=200$mm. Right panel,
     theoretical solution using the results of Ref.\ \cite{Gray06}
     with $Pe=14$.}
 \label{fig:densityLessFilled}
 \end{figure*}

The temporal evolution of the concentration is something particularly
difficult to measure experimentally and, to the best of our knowledge,
there is only one realisation \cite{Wiederseiner2011}. Our results
show a strikingly good fit with the theory,
Fig.\ \ref{fig:densityLessFilled}. For the sake of space, we refer the
reader to the original paper for the derivation of the solution
\cite{Gray06} and to the appendix for the explicit formula we
use. Note that for a tapered drum the half volume height is at $z_{50}
= L/\sqrt(2)$, the spatial coordinate of the theoretical solution has
been scaled to account for this.

The results are remarkably similar concerning both the evolution of
the concentration and the spatial distribution. However, we cannot fit
both the temporal evolution and the steady state solution. We chose to
fit the P\'eclet number $Pe$ (a dimensionless number that quantifies
how large the segregation is compared to diffusive remixing) to
reproduce the evolution of the system rather than the steady
state. Does this mean that the $Pe$ in the tapered drum changes with
time? Answering this question goes beyond the scope of this paper.

It must be noted that we are comparing the results of only
one realisation, and our data points are not averaged on time, which
would decrease the noise and give a better fit, but we prefer to show
the raw results to highlight the quality of the data that is possible
to obtain with this set-up. We confirmed that the same qualitative
behaviour holds for different angles and drum lengths (data not
shown).

Finally, it must be noted that Gray \& Chugonov's theory can be used
to explain the centre of mass movement reported in
\cite{Staron2014,Thornton2011}, together with our results. Consider
the solution for the concentration of small particles in the chute,
$\phi(x,t)$. To obtain the centre of mass of the small particles one
integrates the concentration times the position, $z_{cm}(t) = \int_0^1
z ~\phi(z,t)~ dz~,$ which can be solved numerically in a very
straightforward way. It is easy to verify that the predicted solution
shows an exponential decay for small values of $Pe \sim 4$, but seems
to show two different slopes for larger values of it . In this way,
$Pe$ can be measured not only for the steady state solution, as done
in \cite{Thornton2011} but from the evolution of the system. The
results presented here indicate that both may be different.

A detailed study of $Pe$ for different system parameters is beyond the
scope of this paper. Our objective being to show how well the theory
describe the evolution of the system and how the tapered drum could
become a trivial experimental set-up where to measure the P\'eclet
number. A detailed study of this should be hopefully a subject to be
addressed both experimentally and theoretically in the future.

\section{Conclusions}
We have studied the size-driven segregation of bi-disperse mixtures in
tapered rotating drums. The relation between velocity field, free
surface, and the segregation pattern was studied. The results confirm
that level differences on the free surface create axial flows, which
in turn cause axial segregation. For the first time we report total
axial segregation in this particular geometry and we studied its
dependence on different parameters; in particular, we showed that the
axial segregation speeds up with the angle of the drum.

We have shown that tapered drums can be used to determine the P\'eclet
number used in previous theories from a straightforward experimental
set-up. This has been done before but the experiments are very
difficult to set up ``with both well controlled initial conditions and
a steady uniform bulk flow field'' \cite{Wiederseiner2011}. And
indeed, to the best of our knowledge, the results from Wiederseiner et
al. are the only ones that give the P\'eclet number experimentally. On
the contrary, to obtain well controlled results on a tapered drum is
extremely easy, for example, using the same set-up that motivated this
study \cite{Kawaguchi2006}.


Besides these theoretical implications, the results presented here
have direct applications in the separation and sorting of granulates
in areas ranging from powder handling in the pharmaceutical industry
to ore milling in the mining industry and shed light on the important
role of geometry in segregation processes, not only for rotating drums
but everywhere tridimensional flows appear. 

\section*{ACKNOWLEDGMENTS}
This work was supported by the IPROCOM Marie Curie initial training
network, funded through the People Programme (Marie Curie Actions) of
the European Union's Seventh Framework Programme FP7/2007-2013/ under
REA grant agreement No. 316555. We would like to thank J.\ Kihnast, N.\ Rivas,
W.\ den Otter, S.\ Luding and A.R.\ Thornton for fruitful discussions,
and the kind hospitality of A.\ Michrafy at the Ecole de Mines, Albi,
France, where part of work was realised. The simulations performed
for this paper were undertaken in Mercury-DPM. It is primarily
developed by T.\ Weinhart, A.R.\ Thornton and D.\ Krijgsman at the
University of Twente.

\section*{Appendix I}
The general solution for the concentration of small particles can be
written as

\[ \phi = \frac{1}{2}\left( 1- \frac{2}{\omega}\frac{\partial\omega}{\partial \xi}\right),\]
where $\omega$ is decomposed onto a steady and a time-varying part
$\omega = \omega_s + \omega_\tau$. Their analytical forms is given by
\[ \omega_s = \chi_0(\xi_0)\frac{\sinh(\xi/2)}{\sinh(\xi_0/2)} - \frac{\sinh((\xi- \xi_0)/2)}{\sinh(\xi_0/2)}~,\]
for the steady solution, while the temporal part is given by
\[\omega_\tau = \sum_{n=1}^\infty A_n \exp\left( -(\frac{n^2\pi^2}{\xi_0^2} + \frac{1}{4})\tau\right)\sin(\frac{n\pi\xi}{\xi_0})~.\]
where the transformed time and space variables are given by $\tau =
(S_r^2/D_r)t$ and $\xi = (S_r/D_r) z$ and $\xi_0 = Pe$. For an
equal volume mixture, as in our case, $\chi_0(\xi_0) = 1$. Using
Eq.\ (5.46) from \cite{Gray06} for our mean concentration $\phi_m =
0.5$, we find that the coefficients in the Fourier sine series are
simplified into

\[A_n = \frac{2Pe^2}{n\pi(Pe^2+4n^2\pi^2)} \left(1 - (-1)^n \right) .\]
In our solution, we take only the first hundred terms of the series.

\bibliography{new200409,granulates,biblio}

\end{document}